\documentclass[usenatbib]{mn2e}
\usepackage{float}
\usepackage{epsfig}
\usepackage{amssymb}
\usepackage{amsmath}
\usepackage{color}
\usepackage[usenames,dvipsnames,svgnames,table]{xcolor}

\newcommand{\bn}{\begin{enumerate}}
\newcommand{\en}{\end{enumerate}}
\newcommand{\bi}{\begin{itemize}}
\newcommand{\ei}{\end{itemize}}

\newcommand{\Msun}{M_\odot}

\newcommand{\himpc}{h^{-1} {\rm Mpc}}
\newcommand{\hikpc}{h^{-1} {\rm kpc}}

\newcommand{\Muv}{M_{\rm uv}}
\newcommand{\Muvth}{M_{\rm uv}^{\rm th}}
\newcommand{\DCs}{DC_{\rm SFH}}
\newcommand{\DCm}{DC_{M_{\rm uv}}}
\newcommand{\Gyr}{\rm Gyr}
\newcommand{\Myr}{\rm Myr}
\newcommand{\Lam}{\Lambda}
\newcommand{\tduty}{t_{\rm duty}}
\newcommand{\Ebv}{E_{B-V}}
\newcommand{\Msunr}{\Msun\,{\rm yr}^{-1}}

\newcommand{\apj}{ApJ}
\newcommand{\aap}{A\&A}
\newcommand{\apjl}{ApJL}
\newcommand{\apjs}{ApJS}
\newcommand{\mnras}{MNRAS}
\newcommand{\aj}{AJ}
\newcommand{\araa}{ARA\&A}

\title[Duty Cycle]{Duty Cycle and the Increasing Star Formation History of $z\geq 6$ Galaxies }

\author[Jaacks et al.]{Jason Jaacks$^1$\thanks{Email: jaacksj@physics.unlv.edu}, Kentaro Nagamine$^1$\thanks{Visiting Scientist. Kavli Institute for the Physics and Mathematics for the Universe (IPMU), University of Tokyo, 5-1-5 Kashiwanoha
Kashiwa, 277-8583, Japan}, 
Jun-Hwan Choi$^2$ \vspace{0.3cm}\\
$^1$ Department of Physics \& Astronomy, University of Nevada, Las Vegas, 4505 S. Maryland Pkwy, Las Vegas, NV, 89154-4002, USA \\
$^2$ Department of Physics \& Astronomy, University of Kentucky, Lexington, KY 40506, USA.
}

\begin{document}
\maketitle


\begin{abstract}
We examine the duty cycle and the history of star formation (SFH) for high-redshift galaxies at $z\ge 6$ using cosmological hydrodynamic simulations.  We find that, even though individual galaxies have bursty SFH,  the averaged SFH between $z\sim 15$ to $z=6$ can be characterized well by either an exponentially increasing functional form with characteristic time-scales of 70\,Myr to 200\,Myr for galaxies with stellar masses $M_s \sim 10^6 \Msun$ to $>10^{10} \Msun$ respectively, or by a simple power-law form which exhibits a similar mass dependent time-scales.  Using the SFH of individual galaxies, we measure the duty cycle of star formation ($\DCs$); i.e., the fraction of time a galaxy of a particular mass spends above a star formation rate (SFR) threshold which would make it observable to the Hubble Space Telescope ({\it HST}) during a given epoch.  We also examine the fraction of galaxies at a given redshift that are brighter than a rest-frame UV magnitude ($\Muv\sim -18$), which is sufficient enough to make them observable ($\DCm$).  We find that both $\DCs$ and $\DCm$ make a sharp transition from zero (for galaxies with $M_s \lesssim 10^7 \Msun$) to unity (for $M_s > 10^9 \Msun$).  The measured duty cycle is also manifested in the intrinsic scatter in the $M_s - SFR$ relationship ($\sim1$ dex) and $M_s - \Muv$ relationship ($\Delta \Muv \sim \pm 1$\,mag).  We provide analytic fits to the $DC$ as a function of $M_s$ using a sigmoid function which can be used to correct for catalogue incompleteness.  We consider the effects of duty cycle to the observational estimate of galaxy stellar mass functions (GSMF) and the star formation rate density (SFRD), and find that it results in a much shallower low-mass end slopes of the GSMF and a reduction of $\gtrsim 70$\% of our intrinsic SFRD, making our simulation results more compatible with observational estimates.  
\end{abstract}



\section{Introduction}
\label{sec:intro}

Cosmological studies of structure formation have made tremendous progress over the past decade. 
At low redshift, large galaxy surveys such as the SDSS and 2dF have measured the galaxy distribution with high accuracy \citep[e.g.,][]{Tegmark.etal:04, Cole.etal:05}.   Other cosmological probes, such as the distances to the Type Ia supernovae \citep{Riess.etal:96,  Permutter.etal:99}, cluster abundance \citep{Bahcall.etal:97}, temperature anisotropy of cosmic microwave background radiation \citep{Spergel.etal:00, Komatsu.etal:11} are also employed. 
A combination of these observational results all point to a leading theoretical model of our Universe that is dominated by cold dark matter (CDM) and dark energy ($\Lam$).  Our goal is to draw a self-consistent picture of galaxy formation and evolution under the framework of $\Lam$CDM model. 

The observations of local galaxies often show that the SFH of local galaxies can be fitted well with a exponentially declining functional form \citep{Gallagher.etal:84, Sandage:86, Kennicutt:98}.  However, in the $\Lam$CDM model, small objects form first, and they later merge into more massive systems. Therefore one would expect that there must be an early phase of galaxy formation when stars are vigorously formed in small dark matter halos, and larger galaxies are being assembled for the first time with an increasing SFH.  The so-called Lilly-Madau diagram \citep{Lilly.etal:96, Madau.etal:96} reflects this rise and fall of cosmic SFH.  The initial version of this diagram had its peak around $z\sim 1$, however, as the observations became deeper and deeper, observers found a large number of fainter galaxies, and soon it was considered to be flat during $z\sim 1-3$ or to even higher redshifts \citep{Steidel.etal:99}.  The redshift frontier was pushed to earlier epochs by the Lyman-break drop-out technique \citep{Steidel.etal:96}, and a larger number of galaxies were detected at $z=3-6$, again showing a decline of cosmic star formation rate density (SFRD) towards higher-$z$ \citep[e.g.,][]{Ouchi.etal:04, Bouwens.etal:04, Giavalisco.etal:04, Bouwens.etal:12}.  

With the advent of the new WFC3 camera onboard the {\it HST}, astronomers have begun to catalog large numbers of candidate star-forming galaxies at $z\ge 6$ \citep[e.g.,][]{Ouchi.etal:09, Bouwens.etal:10a, Wilkins.etal:11a, Yan.etal:10,Oesch.etal:12a}. This is again achieved by the efficient selection of candidates in the color-color plane using the Lyman-break drop-out technique.  However, without a spectroscopic follow-up, physical properties such as stellar mass, SFR, age and metallicity are very difficult to derive accurately. 
The current analysis relies on the SED fitting of a combination of stacked WFC3 rest-frame UV data and Spitzer IR data \citep[e.g.,][]{Labbe.etal:10, Schaerer.deBarros:10, McLure.etal:11, Trenti.etal:11} using population synthesis codes.  
This fit has several free parameters including stellar mass, metallically and age, and it requires an assumption about the SFH of the galaxy.  Often case, the assumptions include an instantaneous burst of star formation followed by a decline, or a burst with a constant SFR.  However the SFH is not well understood for galaxies at $z\ge 6$, and it would be helpful to use a physically motivated models in deriving the age and stellar mass of these high-$z$ galaxies.  As we argued earlier, one would expect an increasing mean SFH at high-$z$, and some researchers have already found such SFH in cosmological hydrodynamic simulations \citep{Finlator.etal:11}, as well as in some observational data \citep{Lee.etal:11,Papovich.etal:11}. 

In our previous work, we have measured the evolution of galaxy luminosity function (LF) and GSMF in our cosmological hydrodynamic simulations \citep{Jaacks.etal:12}.  One of the notable results was the very steep faint-end (low-mass end) slopes of simulated galaxy LF/GSMF, with slopes steeper than $\alpha = -2.0$ at $z\ge 6$ and magnitudes fainter than $\Muv = -19$ and $M_s < 10^9\Msun$.  These simulation results are somewhat incompatible with the current observational estimates \citep{Bouwens.etal:12, Gonzalez.etal:10}, both of whom find less steep faint-ends in the observed LF and GSMF.  However, at the same time we found that the low-mass, star-forming galaxies are necessary to maintain the ionization of intergalactic medium at $z\ge 6$ (see also \citet{Trenti.etal:10, Salvaterra.etal:11, Bouwens.etal:12, Finkelstein.etal:12}), therefore important for the reionization of the Universe.  Thus we realized that it is important to understand the SFH of these high-$z$ galaxies and the duty cycle of star formation in our simulations, which is exactly what we will do in the current paper using the same simulation sets as in \citet{Jaacks.etal:12}.

The discrepancies mentioned above between our simulations and recent observation is less evident when looking at photometric properties (i.e., color-color diagrams, LF), but dramatically more obvious when comparing physical properties (e.g., GSMF, SFRD, $M_s -$SFR relation).  The difficulty of estimating these properties for high-$z$ galaxies is in large part due to the intrinsic scatter of derived physical properties, which can lead to a large scatter in the $M/L$ ratio \citep{Bouwens.etal:11b}.  By examining the specific details of SF in these galaxies, in particular the episodic nature of SF history (i.e., duty cycle), we can better understand the discrepancies between observation and theory.

The idea of using the SF duty cycle as an explanation for the offset between the observed and theoretical GSMF has been demonstrated in previous works.  For example,  \citet{Lee.etal:11} considered a scenario in which the SFHs of lower luminosity galaxies are bursty.  They used the observed $M_s-\Muv$ relation at $z=4$ and it's deviation from the smooth growth model to estimate the duty cycle. 
The model with DC successfully lowers the number density of low-mass objects that would be observed. 
The goal of this paper is to provide the measurement of the duty cycle of star formation of simulated galaxies in our cosmological hydrodynamic simulation without any ad hoc assumptions on gas dynamics. 

This paper is organized as follows.  We briefly describe our simulations in Section\,\ref{sec:sim}. We then present our results of SFH and duty cycle in Section\,\ref{sec:sfh} \& \ref{sec:duty}.  We discuss the implications of our results on our simulated GSMF and SFRD in Section\,\ref{sec:app}, and summarize and discuss in Section\,\ref{sec:con}.


\section{Simulations}
\label{sec:sim}

\begin{table*} 
\begin{center}
\begin{tabular}{cccccc}
\hline
Run  & Box size & $N_{p}$ & $m_{DM}$ & $m_{\rm gas}$ & $\epsilon$  \\ 
& ($\himpc$) & (DM, Gas)  & ($h^{-1} \Msun$) & ($h^{-1} \Msun$) & ($\hikpc$) \\
\hline
N400L10 & $10.0$ & $2 {\times} 400^{3}$ & $9.37{\times} 10^{5}$ & $1.91 {\times} 10^{5}$ & $1.0$  \\
N400L34 & $33.75$ & $2 {\times} 400^{3}$ & $3.60 {\times} 10^{7}$ & $7.34 {\times} 10^{6}$ & $3.38$  \\
N600L100 & $100.0$ & $2 {\times} 600^{3}$ & $2.78 {\times} 10^{8}$ & $5.65 {\times} 10^{7}$ & $4.30$  \\
\hline
\end{tabular}
\caption{Simulation parameters used in this paper. The parameter $N_p$ is the number of gas and dark matter particles; $m_{\rm DM}$ and $m_{\rm gas}$ are the particle masses of dark matter and gas; $\epsilon$ is the comoving gravitational softening length.  }
\label{tbl:Sim}
\end{center}
\end{table*}

We use a modified version of the smoothed particle hydrodynamics (SPH) code GADGET-3 \citep[originally described in][]{Springel:05}.  This modified code includes radiative cooling by H, He, and metals \citep{Choi:09}, heating by a uniform UV background of a modified \citet{Haardt:96} spectrum \citep{Katz:96, Dave:99, Faucher.etal:09},  
an \cite{Eisenstein&Hu:99} initial power spectrum, star formation via the "Pressure model" \citep{Schaye:08, Choi:10a}, supernova feedback, sub-resolution model of multiphase ISM \citep{Springel:03}, and the multicomponent variable velocity (MVV) wind model \citep{Choi:11a}.  Our current simulations do not include AGN feedback.  

We have shown earlier that the metal line cooling enhances star formation across all redshifts by about $10-30$\% \citep{Choi:09}, and that the Pressure SF model suppresses star formation at high-redshift due to a higher threshold density for SF \citep{Choi:10a} with respect to the earlier model by \citet{Springel:03}. \citet{Choi:11a} also showed that the MVV wind model, which is based on the momentum-driven wind, makes the faint-end slope of GSMF slightly shallower compared to the constant velocity galactic wind model of \citet{Springel:03}.  

Simulations are setup with either $2\times 400^3$ or $2\times 600^3$ particles for both gas and dark matter. Multiple runs were made with comoving box sizes of $10h^{-1}$Mpc, $34h^{-1}$Mpc and $100h^{-1}$Mpc to cover a wide range of halo and galaxy masses.  Throughout this work they will be referred to as N400L10, N400L34 and N600L100 runs, respectively. Simulation parameters are summarized in Table~\ref{tbl:Sim}, and the adopted cosmological parameters are based on the WMAP data: ${\Omega_{\rm m}}=0.26$, ${\Omega_{\Lambda}}=0.74$, ${\Omega_{\rm b}}=0.044$, $h=0.72$, $\sigma_{8}=0.80$, $n_{s} =0.96$  \citep{Komatsu:09}.

At each time step of the simulation, the gas particles that exceed the SF threshold density ($n_{\rm th}^{\rm SF}=0.6$\,cm$^{-3}$) are allowed to spawn star particles with the probability consistent with the intended SFR.  Each star particle is tagged by stellar mass, metallicity, and formation time. 
\citet{Choi:10a} have shown that the Pressure SF model with the above $n_{\rm th}^{\rm SF}$ produces favorable results compared to the observed \citet{Kennicutt:98} law, in particular at the low column density end.   Galaxies are then identified by a group-finder algorithm using a simplified variant of the SUBFIND algorithm \citep{Springel:01}.  Collections of star and gas particles are grouped based on the baryonic density field.  Properties such as stellar mass, SFR, metallicity and position, among others are recorded in galaxy property files.  A more detailed description of the group-finder process can be found in \citet{Nagamine:04}.


\subsection{Numerical resolution}

\begin{figure}
\centerline{\includegraphics[width=1.10\columnwidth,angle=0] {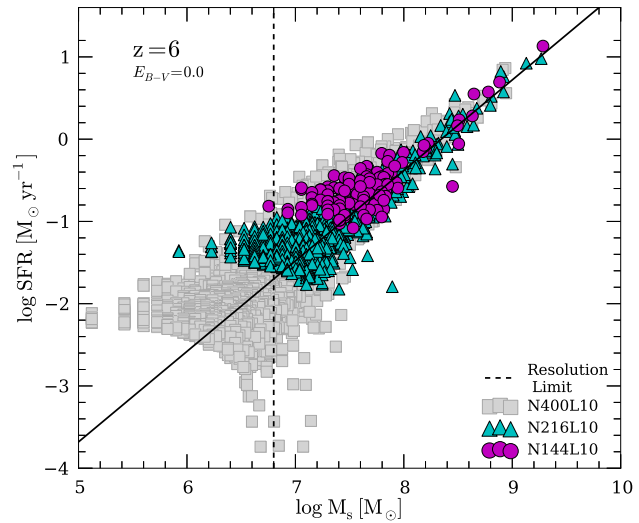}}
\caption{$M_s-SFR$ relationship at $z=6$ from different resolution run, $2\times144^3$ (magenta circles), $2\times216^3$ (cyan triangles) and $2\times 400^3$ (grey squares) all in a $10h^{-1}$ Mpc box size.  The solid black line represent the full dynamic range (Fig.~\ref{fig:muv_sfr}) median fit to this relationship which go as $M_s\propto SFR^{1.10}$.  The vertical dashed black line represents our imposed resolution limit for the N400L10.
}
\label{fig:conv}
\end{figure}

To demonstrate that the simulated $M_s-SFR$ relationship which is used extensively in the work is resolution independent, we present a convergence study in Fig.~\ref{fig:conv}.  Runs with $2\times144^3$ (magenta circles), $2\times216^3$ (cyan triangles) and $2\times 400^3$ (grey squares) all in a comoving $10h^{-1}$ Mpc box size are compared to the full dynamic range median relationship (black solid line, see Fig.~\ref{fig:muv_sfr}) at $z=6$. 
The results from the three runs with different resolution is consistent with the black solid line, suggesting that our result on $M_s-SFR$ relationship is not affected by the resolution. 

The vertical dashed black line represents our imposed resolution limit for the N400L10 run and corresponds to a galaxy with approximately 66 star particles.  Galaxies below this resolution limit are excluded from consideration when calculating the $M_s-SFR$, $M_s-\Muv$, GSMF, \& SFRD.  
\citet{Jaacks.etal:12} have shown that the gas particle mass of the N400L10 run is lower than the typical Jeans mass at $z=6$ by a factor of  $\approx 1-100$, but the run does not satisfy the \citet{Bate&Burkert:97} mass resolution criteria.  Therefore our simulations are not resolving the collapse of star-forming molecular clouds directly, which is why we need to employ the sub-resolution multiphase model for star formation in our current simulations.

\section{Results}

\subsection{Star Formation History}
\label{sec:sfh}
\begin{figure}
\centerline{\includegraphics[width=1.10\columnwidth,angle=0] {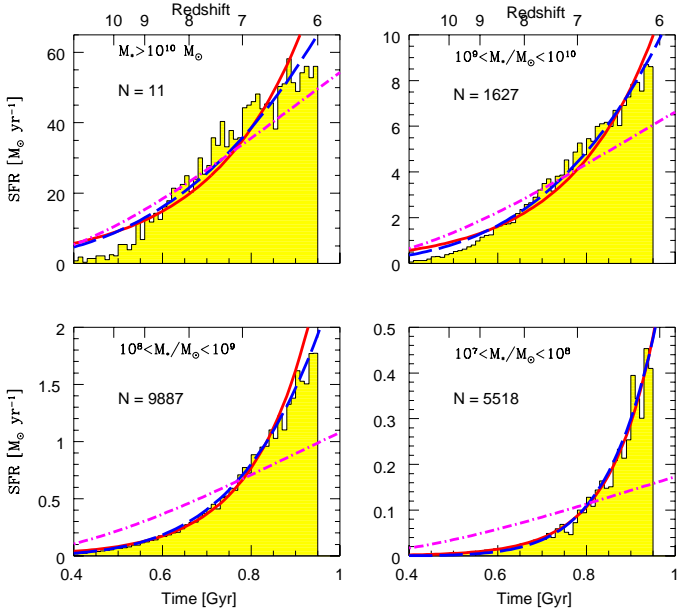}}
\caption{
Average star formation histories for different groups of galaxies divided by their stellar mass at $z=6$ with time bins of $\Delta t=10$\,Myr.  The red solid lines show the least square fit to an exponential function, and the blue dashed lines show the power-law fit.  The magenta dashed-dotted line represents the polynomial fit to $SFR/\left<SFR\right>$ found by \citet{Finlator.etal:11}.  To make a valid comparison the value of $\left<SFR\right>$ is computed for each panel and used to show the function.  The total number ($N$) of galaxies found in each mass group is also shown in the plot.  The best-fit parameters can be found in Tables~\ref{tbl:fit_exp} \& \ref{tbl:fit_power}.
}
\label{fig:sfh6}
\end{figure}

Fig.~\ref{fig:sfh6} shows the mean SFH for all galaxies found in our three runs, N400L10, N400L34 and N600L100.  
It is computed as follows.  First, the SFH for each identified galaxy is calculated by counting the number of star particles that were created in each time bin of $\Delta t = 10$\,Myr.  Galaxies are then grouped by their stellar masses at $z=6$, and the mean SFH is plotted for each stellar mass group.  Note that the SFH at higher redshifts includes all the star formation in all progenitor galaxies, which may have had lower stellar masses than the final mass at $z=6$. 

To quantify the mean SFH, we first perform a least square fit to an exponential function (solid red line in Fig.~\ref{fig:sfh6}),
\begin{equation}
\label{eq:exp}
SFR=A\ \exp \left (  \frac{t- t_{z6}}{\tau_E} \right ) \ [\Msun\ \Gyr^{-1}], 
\end{equation} 
where $t$ is the cosmic time since the Big Bang, $A$ is the normalization at $z=6$ in units of $\Msun\,\Gyr^{-1}$, $t_{z6}$ is the cosmic time at $z=6$, and $\tau_E$ is the characteristic time-scale. 
We vary the values of $A$ and $\tau_E$ to obtain the best-fitting function. 
We find that the mean SFH can be described fairly well by the increasing exponential function for all stellar mass groups with a $\tau_E$ value ranging from $70$ to $200$\,Myrs (see Table~\ref{tbl:fit_exp} for the best-fit parameters).  This procedure was repeated for galaxies at $z=7,8,9$ with similar results (not shown here).

We also perform the same least square fitting procedure to a power-law function (blue dashed), 
\begin{equation}
\label{eq:pwr}
\frac{SFR}{\left < SFR \right >}= \left ( \frac{t}{\tau_P} \right )  ^{\alpha} \ [\Msun\ \Gyr^{-1}],
\end{equation}
where $\left<SFR\right>$ is the mean SFR calculated between $t=0$ and $t=t_{z6}$, $\tau_P$ is the characteristic time-scale when $SFR=\left<SFR\right>$,  and $\alpha$ is the power index.  See Table~\ref{tbl:fit_power} for the best-fitting values.  

In the last column of Tables~\ref{tbl:fit_exp} and \ref{tbl:fit_power}, the values of correlation coefficients are listed as a measure of goodness of fit.  A value closer to unity indicates a better fit. 
In most cases the power-law function does a slightly better job of fitting the shape of the SFH, but they both provide a very good fit. 

\citet{Hernquist.etal:03} found that the cosmic SFH in their cosmological SPH simulations can be fitted well by $\dot{\rho}_\star \propto \exp (-z/3)$ at high-redshifts, and speculated that this exponential behavior is related to the growth of dark matter halo mass function. Our results support their argument, and we also suggest that the early star formation history is driven by the gravitational instability of dark matter structure, and it is natural to expect an exponential or increasing power-law behavior of SFH as a function of cosmic time.

\begin{table}
\begin{center}
\begin{tabular}{cccc}
\hline
Mass Range & A  & ${\tau_E}$ & $r^2$ \\
($h^{-1} M_{\odot}$) & ($\Msun$\,Gyr$^{-1}$) & (\Gyr)& \\ 
\hline
$M_s>10^{10} M_{\odot}$ & $76.70$ & $0.21$ & $0.9212$ \\
$10^{9}<M_s /M_{\odot}<10^{10}$ & $12.30$ & $0.16$ & $0.9700$\\
$10^{8}<M_s /M_{\odot}<10^{9}$ & $3.17$ & $0.11$ & $0.9976$\\
$10^{7}<M_s /M_{\odot}<10^{8}$ & $0.42$ & $0.10$ & $0.9990$\\
\hline
\end{tabular}
\caption{Best-fitting parameter values to the exponential function (Eq.\,\ref{eq:exp}) for galaxies in different stellar mass ranges, obtained by the least squares fit of the star formation histories at $z\ge 6$.  The value of $A$ is the normalization constant, $\tau_E$ is the characteristic exponential time-scale, and $r^2$ is the correlation coefficient. 
}
\label{tbl:fit_exp}
\end{center}
\end{table}

\begin{table}
\begin{center}
\begin{tabular}{ccccc}
\hline
Mass Range & $\langle SFR \rangle$  & ${\tau_P}$ & $\alpha$ & $r^2$  \\
($h^{-1} M_{\odot}$) & ($M_\odot\ \Gyr^{-1}$) & (\Gyr)& \\ 
\hline
$M_{\star}>10^{10} M_{\odot}$ & $15.24$ & $0.59$ & $3.05$ & $0.9592$\\
$10^{9}<M_s /M_{\odot}<10^{10}$ & $1.88$ & $0.62$ & $3.74$ & $0.9886$\\
$10^{8}<M_s /M_{\odot}<10^{9}$ & $0.30$ & $0.66$ & $5.12$ & $0.9949$\\
$10^{7}<M_s /M_{\odot}<10^{8}$ & $0.05$ & $0.73$ & $8.55$ & $0.9748$\\
\hline
\end{tabular}
\caption{Same as Table~\ref{tbl:fit_exp}, but for the power-law function (Eq.\,\ref{eq:pwr}).  The SFR takes the value of $\langle SFR \rangle$ at $t=\tau_P$, and $\alpha$ is the power-index of the power-law function. 
}
\label{tbl:fit_power}
\end{center}
\end{table}

Included in Fig.~\ref{fig:sfh6} is a comparison to a polynomial fit of the average SFH found by \citet{Finlator.etal:11}, 
which is indicated by the magenta dashed-dotted line.  Using the G{\small ADGET-2} SPH code,  they also found increasing SFH, and argued that the shape of SFH to be mass invariant and only changes by a scale factor through different mass bins, with the following functional form:
\begin{equation}
\log\left(\frac{SFR}{\left<SFR\right>}\right)\propto 0.2-3.2t+10.5t^5-3.9t^3
\end{equation}
for $t>2.7$ and equal to zero elsewhere, where $t$ is the age of the Universe in $\Gyr$.
One can see that our simulation results do not agree well with this functional form. 
In contrast, in our simulations we find that both the exponential and power-law fits to have an evolving $\tau$ with varying galaxy stellar mass.  The source of this difference might be the different models of star formation and feedback in the two codes. 

\begin{figure}
\begin{center}
\includegraphics[scale=0.4,angle=0] {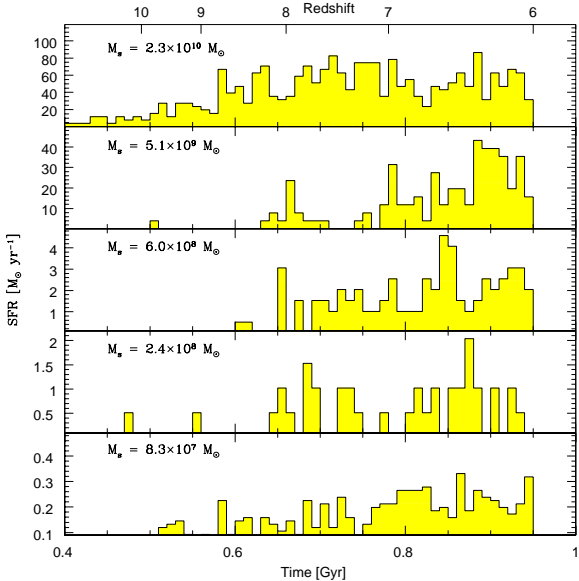}
\caption{
Star formation histories of individual galaxies with different stellar masses as indicated in each panel taken from the N400L34 run, 
with time bins of $\Delta t=10$\,Myr.
On average, they all increase gradually, but individual galaxies show intermittent bursts of star formation. 
}
\label{fig:sfh2}
\end{center}
\end{figure}

In addition to the mean SFH, we also examined the SFH of individual galaxies as shown in Fig.~\ref{fig:sfh2}.
It shows that, although the mean SFH may be smoothly rising, the SFH of the individual galaxies are dominated by intermittent bursts throughout their evolution up to $z=6$.  These burst are thought to be the result of a combination of galaxy merger events \citep{Mihos.etal:96,Kennicutt:98,Khochfar.etal:11} and cold mode accretion onto the galaxies \citep{keres.etal:05}.  Note that the SFHs shown in Fig.~\ref{fig:sfh2} include all the star formation in all progenitors that end up in the galaxy at $z=6$.  We now quantify the bursty nature of star formation in terms of duty cycle in the next section.

\subsection{Duty Cycle}
\label{sec:duty}

The definition of duty cycle ($DC$) of star formation can be somewhat ambiguous. 
In this work we define it in two different ways.  The first, $DC_{\rm SFH}$ is based on the SFH of a galaxy, and is characterized by the fractional time that a galaxy spends above a threshold SFR which is sufficient to make it observable. The second, $DC_{\Muv}$ examines the number fraction of galaxies (in a mass bin at a given instant) that are brighter than the threshold rest-frame UV magnitude $\Muvth$, making it observable. 

\begin{figure}
\centerline{\includegraphics[width=1.10\columnwidth,angle=0] {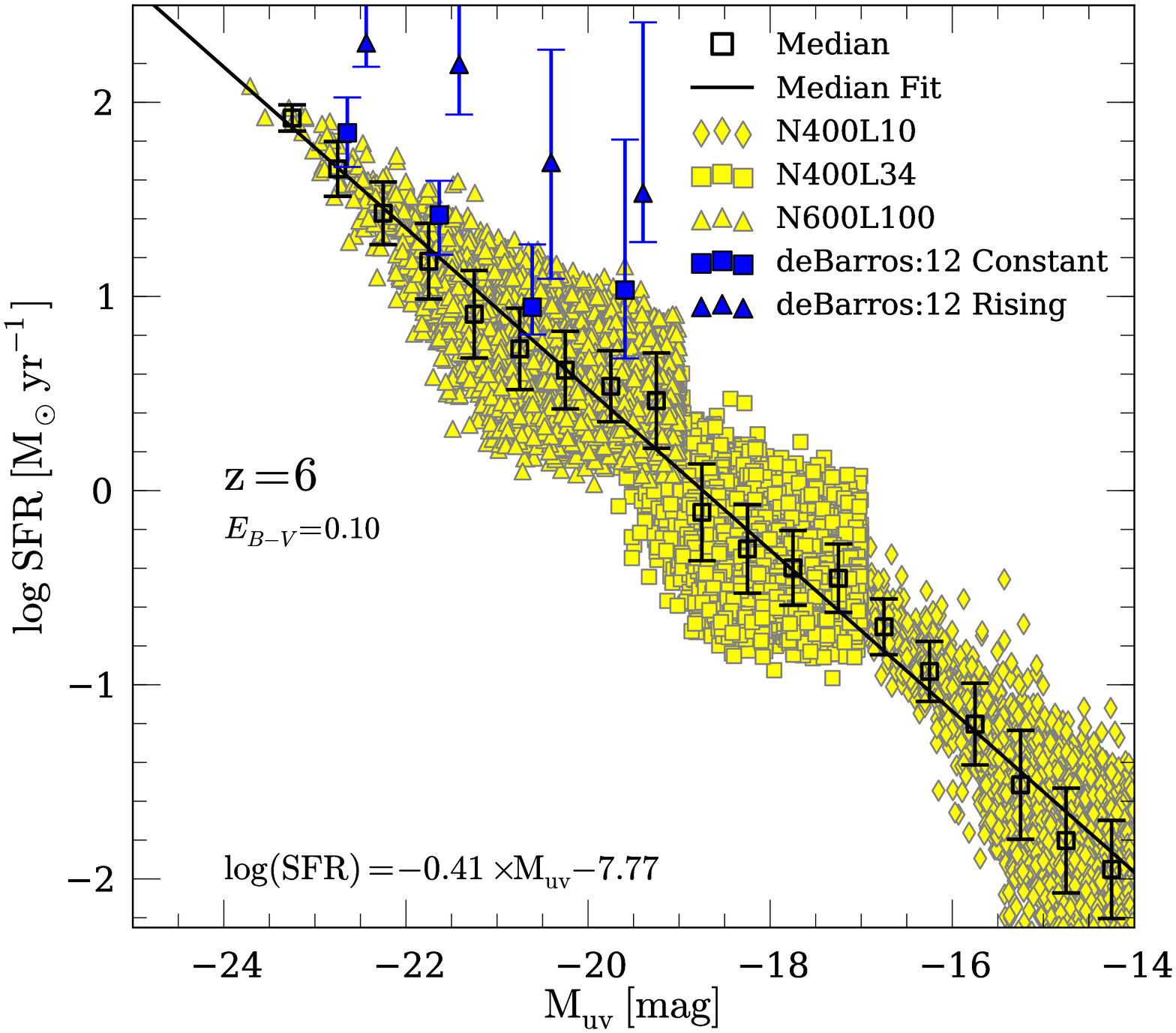}}
\caption{Relationship between $\Muv - SFR$ at $z=6$ with $\Ebv =0.10$.  Yellow squares, triangles and diamonds represent N400L34, N600L10 and N400L10, respectively.  The black open triangles are the median found in each bin with error bars representing one standard deviation.  The solid black line is the least square fit to the median data points.  The blue triangles and squares represent observations from \citet{deBarros.etal:12} where the triangles are based on SED fits which include rising SFH with nebular emission lines and Ly$\alpha$ and the squares are based on SEDs with constant SFH with nebular emission lines and Ly$\alpha$.}
\label{fig:muv_sfr}
\end{figure}

The SFR threshold ($SFR_{\rm th}$) used to determine $DC_{\rm SFH}$ is obtained by converting the HST's magnitude limit of  $M_{\rm AB, lim} \approx 29$ to the rest-frame $\Muvth$ via the standard formula,
\begin{equation}
\label{eq:muv}
\Muvth = M_{\rm AB, lim}-5\log \left( \frac{d_L}{10\mathrm{pc}}\right)+2.5\log\left(1+z\right).
\end{equation}
At $z=6,7,8$, it gives $\Muvth \approx -17.74, -17.98, -18.19$, respectively.  
Then, $\Muvth$ can be converted to $SFR_{\rm th}$ via the relationship between $\Muv$ and SFR of galaxies found in our simulations which goes on average as $SFR \propto \Muv^{-0.40}$ (Fig. \ref{fig:muv_sfr}). This results in $SFR_{\rm th}=0.50\,\Msun\,{\rm yr}^{-1}$ with no consideration for dust extinction, and $SFR_{\rm th}=1.20\,\Msun\,{\rm yr}^{-1}$ when including $E_{B-V}=0.10$.  
This value of  $E_{B-V}$ is chosen to be consistent with the value used to match the observed rest-frame UV luminosity function (LF) in our previous work \citep{Jaacks.etal:12}, and it is centered between the following two recent observations: 
\citet{Bouwens.etal:11b} argued for little to no extinction at the faint-end of the LF at $z=6$, whereas  
\citet{Willott.etal:12} found a best-fit value of $A_V=0.75$, which corresponds to $E_{B-V}\sim 0.19$ assuming $R_V=4.05$ \citep{Calzetti.etal:00} at the bright-end of the UV LF at $z=6$.  This moderate amount of extinction is also consistent with the estimates by \citet{Schaerer.deBarros:10,deBarros.etal:12,Schaerer.etal:12} who included nebular emission lines in their SED fits.  Since the conversion between $\Muvth$ and $SFR_{\rm th}$ depends on the relationship between $\Muv$ and $SFR$ in the simulation, the value of $SFR_{\rm th}$ depends on the assumed value of $E_{B-V}$. 

In Fig. \ref{fig:muv_sfr} we plot uncorrected observational estimates based on SED fits found in \citet{deBarros.etal:12}.  Although blue triangles and squares are median points from the same data set they are derived through SED fits which make different assumptions regarding SFH.  The blue triangles are calculated using a rising SFH, whereas the squares are calculated using a constant SFH.  Both include nebular emission lines and Ly$\alpha$.  At first look it would seem that our data (yellow diamonds, squares and triangles) have better agreement with the blue squares which assume a constant SFH.  This however is slightly misleading due to the fact that \citep{deBarros.etal:12} find that when using the rising SFH on average a high extinction is also obtained.  Therefore this comparison is completely dependent upon our included amount of extinction, i.e. if we were to include more extinction we would have better agreement with the blue triangles.  This is further justification to explore a range of extinction as we do in this work.

Recent $z\geq6$ publications \citep{Bouwens.etal:12,Smit.etal:12} suggest a $\Muv$-dependent extinction based on an observed relationship between $\Muv$ and the UV-continuum slope ($\beta$).  While these results are interesting, we have chosen not to adopt this model, and instead apply constant extinction in an effort to show a range of $\Ebv$ values and their impact systematically.  Any results obtained from using a $\Muv$-dependent model of extinction with fall within the range presented in this work.

We find that the size of time bins $\Delta t_{\rm duty}$ for computing SFH makes a difference in determining the value of DC. We determine its value based on the mass resolution of each simulation and $SFR_{\rm th}$, i.e., $\Delta t_{\rm duty} = m_{\rm star} / SFR_{\rm th}$, where $m_{\rm star}$ is the mass of star particles in the simulation, which is taken to be the half of gas particle mass in the simulation setup.  This method insures that the simulation is able to satisfy the threshold SFR if one star particle forms in any given time bin and allow for the comparison between runs with different mass resolutions.  

Based on the above definition, we adopt time bins of $\Delta \tduty = 10$\,Myr and 0.27\,Myr for the N400L34 and N400L10 runs, respectively, for no extinction, and $\Delta \tduty = 4$\,Myr and 0.22\,Myr for the case with $E_{B-V}=0.10$.  The system time steps of these simulations are typically $\Delta t_{\rm sys} = 10^3 - 10^4$\,yrs at $z= 6-7$, therefore we have at least 10 system time steps in each $\Delta \tduty$, providing reasonable time resolution in each $\Delta \tduty$. 

\begin{figure}
\begin{center}
\includegraphics[scale=0.30,angle=0] {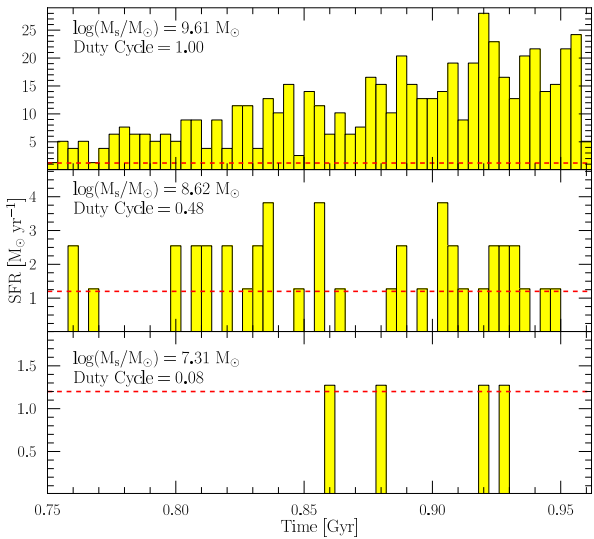}
\caption{
Examples of three different SFH computed with $\Delta \tduty=4\ \Myr$ from our N400L34 run between $z=6-7$ and corresponding $\DCs$ for galaxies of varying masses.  The threshold of SFR$_{\rm th} = 1.20\, \Msun\, {\rm yr}^{-1}$ used to calculate $\DCs$ is shown as the red dashed line. 
}
\label{fig:sfh_thresh}
\end{center}
\end{figure}

Three example SFHs between $z=6-7$ are shown in Fig.~\ref{fig:sfh_thresh} for galaxies in the N400L34 run with stellar masses ranging $\approx 10^7-10^{10} \Msun$ and $\DCs=0.08, 0.48, 1.00$. This figure illustrates the method used to calculate $\DCs$, which is the fraction of time spent above $SFR_{\rm th}=1.20\, \Msun$\,yr$^{-1}$ (red dashed line).

\subsubsection{Duty cycle vs. Galaxy Stellar Mass}
\begin{figure*}
\begin{center}
\includegraphics[width=1.8\columnwidth,angle=0] {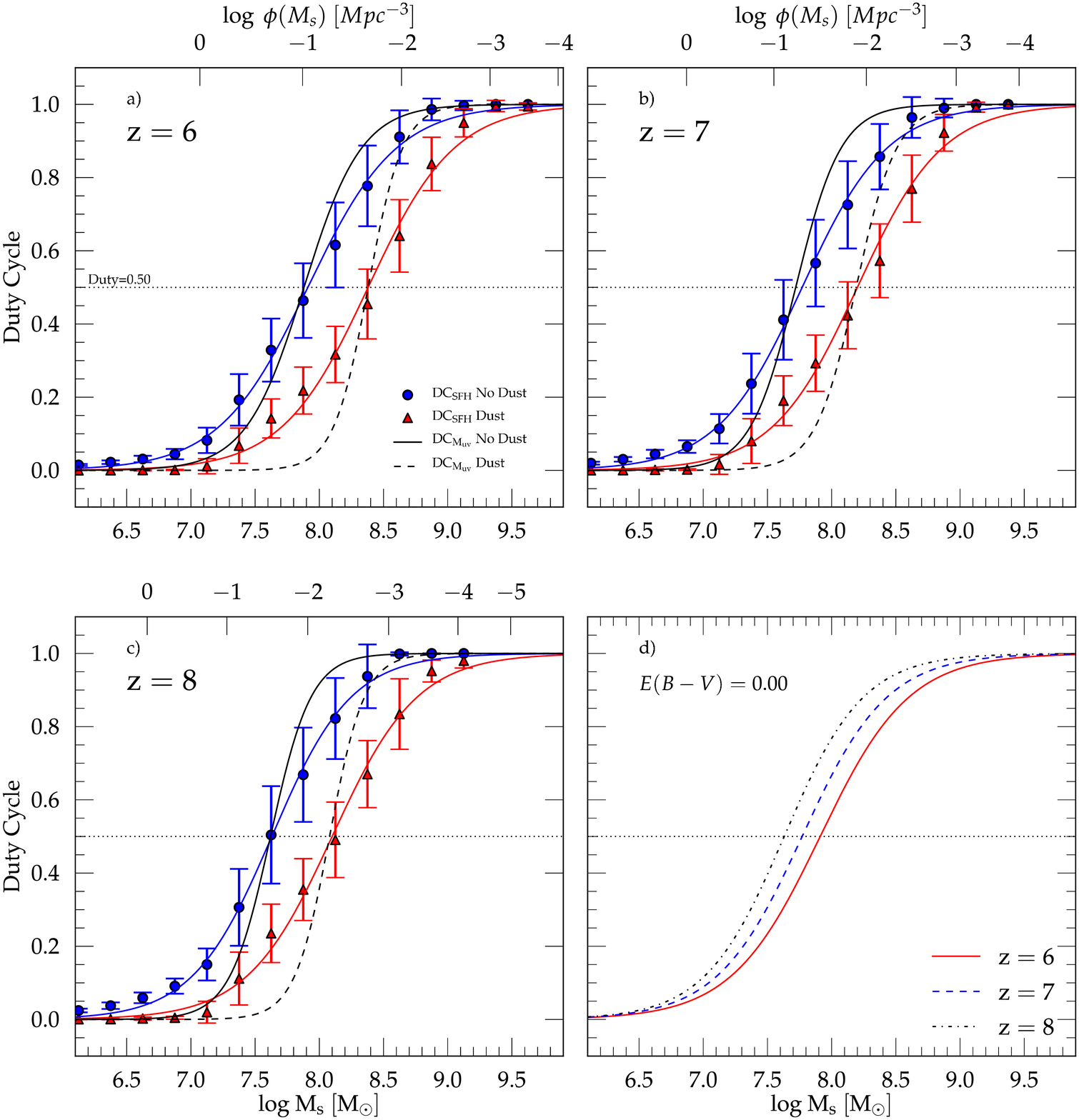}
\caption{Duty cycles ($\DCs$ and $\DCm$) for galaxies in the N400L34 and N400L10 runs at $z=6,7,8$.  Each method is calculated with (red triangles, dashed line) and without dust extinction (blue circles, solid line).  The bottom right panel shows the redshift evolution of $\DCs$ without dust extinction.  The dotted line has been added to indicate $DC=0.50$ to aid comparison, and error bars represent one standard deviation in each mass bin.  The top axes of panels ($a$), ($b$) and ($c$) show the logarithm of comoving number density  ($\log \phi(M_s)$) of objects at a given mass taken from Schechter fits to the intrinsic GSMF at each redshift found in \citet{Jaacks.etal:12}.  The blue and red lines going through the data points represent the fit to Equation~\ref{eq:sig} for each of the data sets.  Fitting parameter values can be found in Table~\ref{tbl:fit_duty}. 
}
\label{fig:duty}
\end{center}
\end{figure*}
As we described above, we compute both $\DCs$ and $\DCm$ for each galaxy in each simulation. We then make a scatter plot of duty cycle as a function of galaxy stellar mass $M_s$, bin the data in each $\log M_s$ bin, and obtain the median in each bin. 
Figure~\ref{fig:duty}a,b,c show the results of $\DCs$ at $z=6, 7, 8$ with dust (red triangles), without dust (blue circles), and $\DCm$ with dust (dashed black line), without dust (solid black line).  
The top axes of panels ($a$), ($b$) and ($c$) show the number density of objects $\phi(M_s)$ at a given mass taken from the Schechter fits to the GSMF at each redshift found in \citet{Jaacks.etal:12}.

We find that both methods of DC exhibit a characteristic steep transition from $DC=0$  to unity, and pass through $DC=0.50$ at approximately the same mass bin at each redshift.  See Table~\ref{tbl:50duty} for the mass-scales at which the DC crosses the value of 0.50.  This transition can be modeled well by employing a sigmoid function described by
\begin{equation}
DC(M_s)=\left[\exp \left( \dfrac{a-\log (M_s)}{b} \right) +1 \right]^{-1}, 
\label{eq:sig}
\end{equation}
where due to its form the parameter $a$ corresponds to the value of $\log (M_s)$ where $DC=0.50$, and $b$ is a measure of the steepness of the transition; i.e as $b \to 0$ the function approaches a step function, and as $b\to \infty$ the transition becomes infinitely flat.  

Least square fits to Equation~\ref{eq:sig} were performed for each $DC$ at each redshift and the results are shown in Fig.~\ref{fig:duty}a,b,c,d as the blue, red, black and black dashed lines in each panel.  Values of the fit parameters for each model at each redshift can be found in Table~\ref{tbl:fit_duty}.  The values of $a$ parameter is essentially the same as those given in Table~\ref{tbl:50duty}, and the closeness of the $a$ values in the two tables show that the sigmoid function provides a good fit. 

We note that the N600L100 run result was omitted from Fig.~\ref{fig:duty}, because its value of $\Delta t_{duty}$ (=30\,Myr) exceeded the dynamical time of typical systems at these redshifts, which is about 10\,Myr for a galaxy of $M_s = 10^8 \Msun$.  In other words, the N600L100 run has insufficient mass resolution to estimate the duty cycle at these redshifts reliably. 

As shown in Fig.~\ref{fig:duty}d, we find that the transition mass-scale shifts to a lower mass at higher redshift, simply reflecting the fact that the galaxies at higher redshifts are less massive than at lower redshifts. 
Similar evolution is found in $\DCm$, which is not shown here.
This redshift evolution trend might seem counter-intuitive at first, because it suggests that we would see more lower mass objects at $z=8$ with high $DC$s than at $z=6$. However by examining the $DC$ as a function of number density (Fig.~\ref{fig:duty}, top axes) we can clearly see that, for a given $\phi(M_s)$, the $DC$ at $z=6$ is higher than at $z=8$.  For example, for galaxies with a comoving number density of $\phi(M_s)=10^{-1}\ [{\rm Mpc}^{-3}]$, 
our results give $\DCs \approx 0.20, 0.15, 0.08$ at $z=6, 7, 8$, respectively. 

Note that our $\DCs$ definition is similar to the one put forth by \citet{Lee.etal:10}, who defined the $DC$ to be the typical duration of star formation with respect to time span covered by the survey.  Since their work is at lower redshifts than ours, making direct comparisons is not really appropriate.  However, the two results seem to be roughly consistent with each other as they find that galaxies with $M_s \le 10^{8.7} \Msun$ are mostly UV-faint and have $DC = 0.2 - 0.4$ at $z\sim 4-5$. 

Although individual galaxies in our simulations do exhibit bursty SFHs, this is by no means the only scenario in which a galaxy can exhibit a $DC$. It is also possible for galaxies with rising or declining SFHs to have a $DC$.  Take for example the average SFHs broken down by stellar mass presented in Fig.~\ref{fig:sfh6}.  If we use Equation~(\ref{eq:exp}), used to describe our exponential rising mean SFH, and apply the concept of $\DCs$ during the time period between $z=6-7$ we find that the $\DCs \approx 1.0,1.0,0.61,0.0$ for the case with dust and $\DCs \approx 1.0,1.0,1.0,0.0$ for the case without.  These transitions from $\DCs=1.0$ to $\DCs=0.0$ occur in the same mass range as those calculated above.

This suggests that a certain $DC$ value does not correspond to a unique SFH, and the mapping between DC and SFH is certainly not one-to-one.  However we consider that it is unlikely that all galaxies have a smoothly rising SFH with the same functional form, therefore the bursty SFH would be more natural outcome of a hierarchical structure formation under the $\Lambda$CDM paradigm. 


\subsubsection{Future Observations}
\label{future}
\begin{figure}
\begin{center}
\includegraphics[scale=0.36,angle=0] {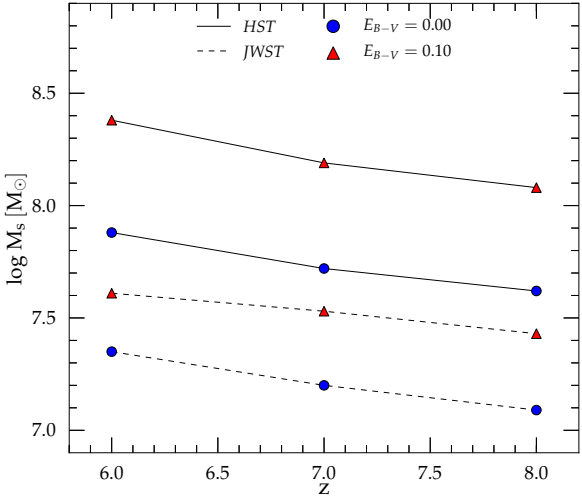}
\caption{The transition stellar mass-scale where $\DCm=0.50$ for both $\Ebv=0.00$ (blue circles) and $\Ebv=0.10$ (red triangles) calculated using {\it HST}'s AB magnitude limit ($M_{AB,lim}\approx 29$) shown by the solid black line and {\it JWST}'s limit ($M_{AB,lim}\approx 31$, dashed black line).}

\label{fig:dz}
\end{center}
\end{figure}

Since both $\DCs$ and $\DCm$ are highly sensitive to the AB magnitude limit of the observing instrument it is an interesting exercise to look a the effect of increase sensitivity.  The James Web Space Telescope ({\it JWST}) is planned to have a magnitude limit of $M_{AB,lim}\approx 31$ \citep{Gardner.etal:06} which, using equation (\ref{eq:muv}), would equate to a $\Muvth \approx-15.75$ at $z=6$.  With this this new threshold, we calculate a SFR$_{\rm th}=0.08\ [\Msunr]$ with $\Ebv=0.0$ and SFR$_{\rm th}=0.30\ [\Msunr]$ with $\Ebv=0.10$, both of which are well below the values calculated above.  Lowering both the $\Muvth$ and SFR$_{\rm th}$ leads to a shift in the $DC$ transition towards lower masses, and this new transition mass-scales can be seen by the dashed black lines in Fig. \ref{fig:dz}. 
For the JWST magnitude limit, the transition galaxy stellar mass-scale is lower by about 0.5 dex and 0.8 dex for $\Ebv=0.0$ and 0.10, respectively.  The JWST should be able to see galaxies down to $M_s \sim 10^7 \Msun$ at $z=8$.

\begin{table}
\begin{center}
\begin{tabular}{cccc}
\hline
$z$ & $Model$ & $E_{B-V}$ & $\log (M_s / \Msun)$  \\
\hline
$6$ & $\DCm$ & $0.00$ &  $7.89$    \\
	& $\DCs$ & $0.00$ &  $7.93$   \\
	 & $\DCm$ & $0.10$ & $8.40$     \\
	 & $\DCs$ & $0.10$ & $8.43$      \\
\hline
$7$ & $\DCm$ & $0.00$ &  $7.72$   \\
	& $\DCs$ & $0.00$ &  $7.77$   \\
	 & $\DCm$ & $0.10$ & $8.20$     \\
	 & $\DCs$ & $0.10$ & $8.25$      \\
\hline
$8$ & $\DCm$ & $0.00$ &  $7.61$    \\
	& $\DCs$ & $0.00$ &  $7.63$   \\
	 & $\DCm$ & $0.10$ & $8.09$    \\
	 & $\DCs$ & $0.10$ & $8.14$     \\
	 
\hline

\end{tabular}
\caption{
The values of $\log (M_s / \Msun)$ at which $\DCm$ and $\DCs$ cross 0.50 at $z=6-8$. 
}
\label{tbl:50duty}
\end{center}
\end{table}

\begin{table}
\begin{center}
\begin{tabular}{ccccc}
\hline
$z$ & $Model$ & $E_{B-V}$ & $a$ &  $b$ \\
\hline
$6$ & $\DCm$ & $0.00$ &  $7.88$ &  $0.225$   \\
	& $\DCs$ & $0.00$ &  $7.91$  & $0.346$   \\
	 & $\DCm$ & $0.10$ & $8.38$  & $0.135$    \\
	 & $\DCs$ & $0.10$ & $8.38$  & $0.336$    \\
\hline
$7$ & $\DCm$ & $0.00$ &  $7.72$ &  $0.177$   \\
	& $\DCs$ & $0.00$ &  $7.77$  & $0.331$   \\
	 & $\DCm$ & $0.10$ & $8.19$  & $0.159$    \\
	 & $\DCs$ & $0.10$ & $8.20$  & $0.331$    \\
\hline
$8$ & $\DCm$ & $0.00$ &  $7.62$ &  $0.158$   \\
	& $\DCs$ & $0.00$ &  $7.63$ &  $0.309$   \\
	 & $\DCm$ & $0.10$ & $8.08$ &  $0.136$   \\
	 & $\DCs$ & $0.10$ & $8.10$  & $0.331$    \\
	 
\hline
\end{tabular}
\caption{
Best fit parameter values of the sigmoid function (Eq.~\ref{eq:sig}) used to characterize $DC$. 
The $E_{B-V}$ column shows the assumed value of extinction in determining the $\Muvth$ and $SFR_{\rm th}$ (see Eq.~\ref{eq:muv}).
}
\label{tbl:fit_duty}
\end{center}
\end{table}


\section{Application of Duty Cycle}
\label{sec:app}
\begin{figure}
\begin{center}
\includegraphics[scale=0.31,angle=0] {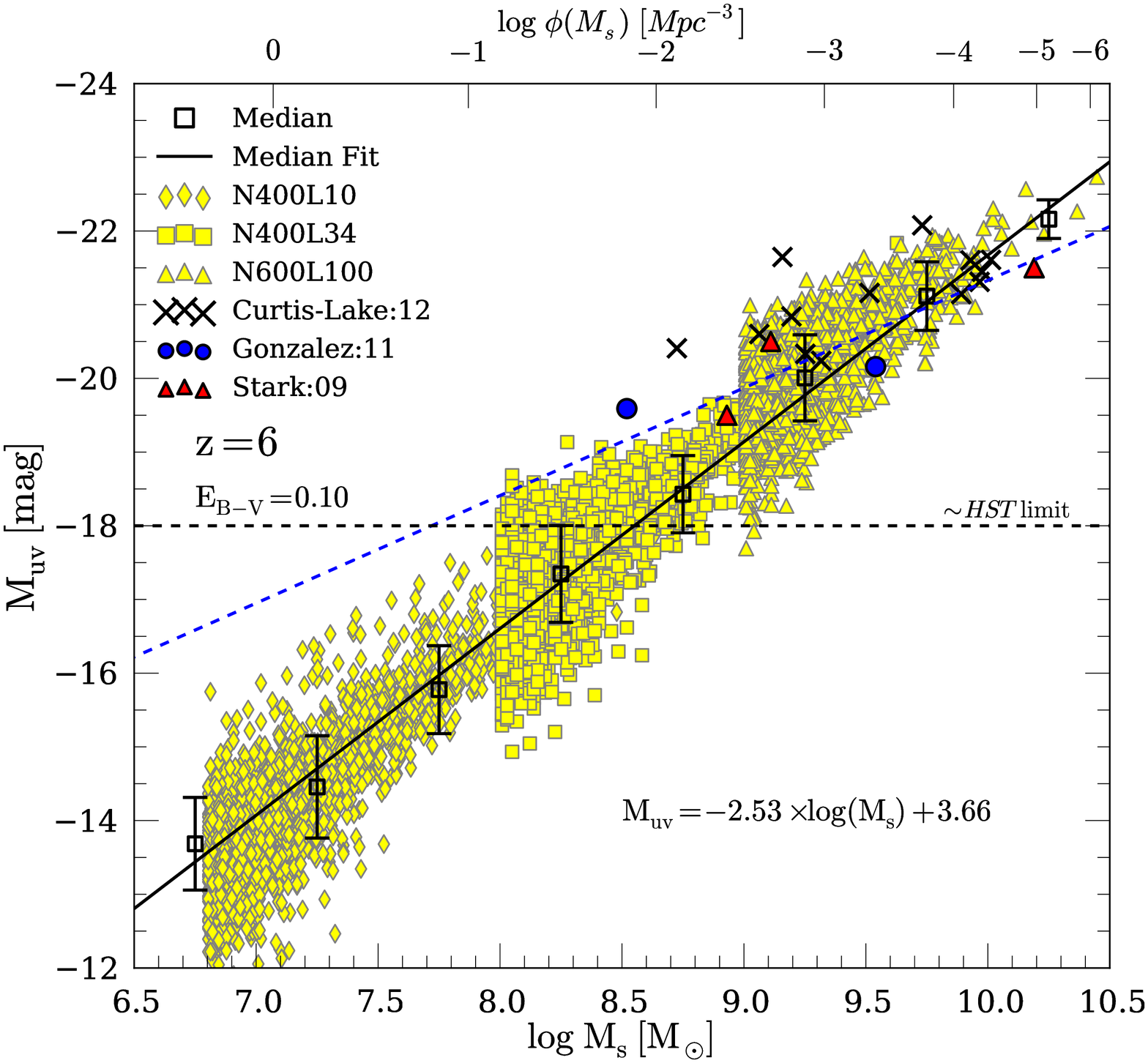}
\caption{Relationship between $Ms-\Muv$  at $z=6$.  Yellow squares, triangles and diamonds represent data from N400L34, N600L100 and N400L10, respectively.  Black open squares are the median found in each bin with the error bars representing one standard deviation.  Solid black line is the least square fit to the median data points.  Blue circles and red triangles represent median points of uncorrected observations taken from \citet{Gonzalez.etal:10,Stark.etal:09} and the black crosses are from recent CANDELS observations found in \citet{CurtisLake.etal:12}.   The blue dashed line is a fit to $z=4$ data points also found in \citet{Gonzalez.etal:10}; note that this is not a fit to the $z=6$ data set presented.  A horizontal dashed black line represents the approximate UV magnitude limit of {\it HST}. 
}
\label{fig:ms_muv}
\end{center}
\end{figure}

\begin{figure}
\begin{center}
\includegraphics[scale=0.325,angle=0] {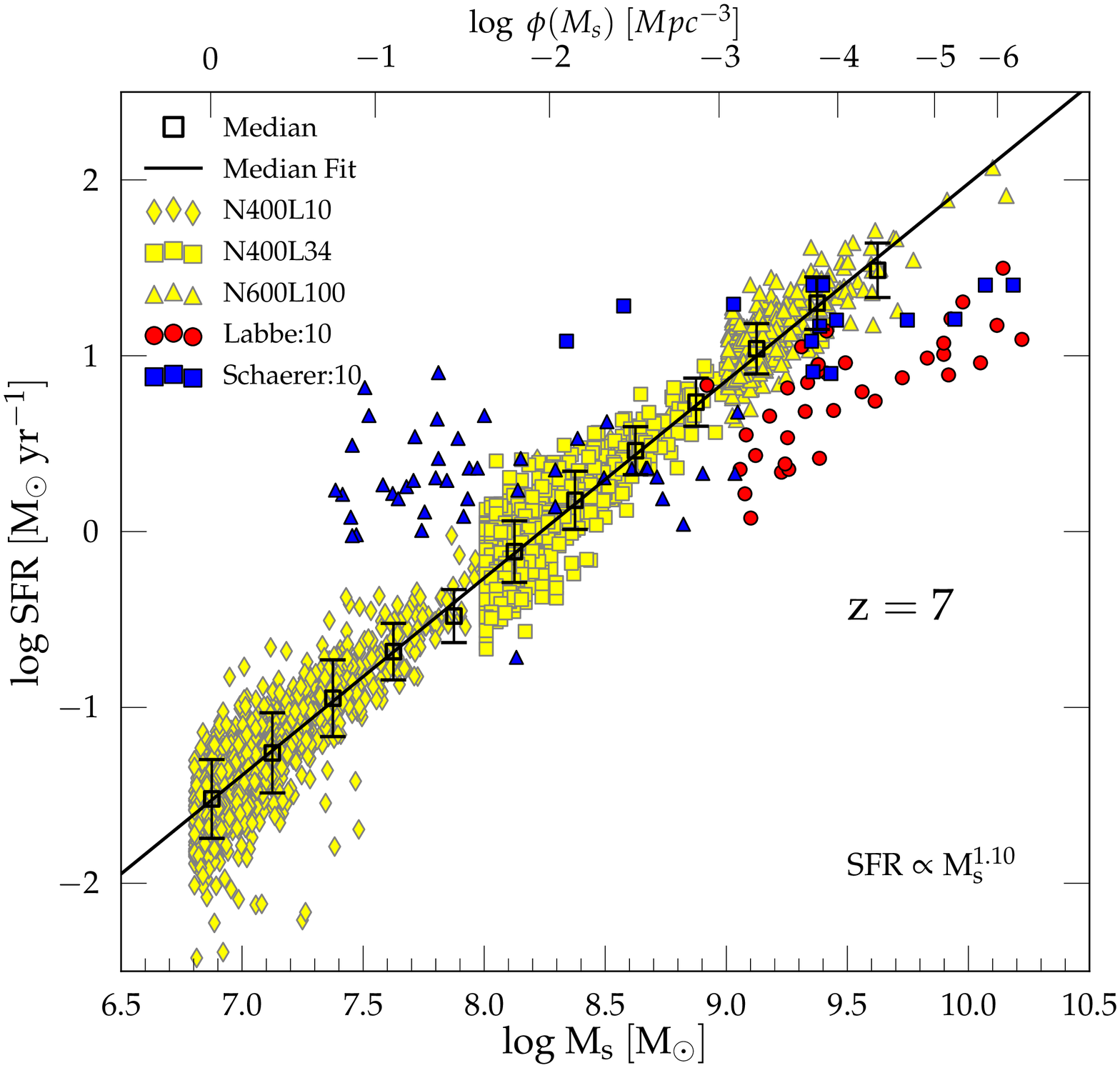}
\caption{Relationship between $M_s-SFR$ at $z=7$.  Yellow squares, triangles and diamonds represent data from N400L34, N600L100 and N400L10, respectively.  Black open squares are the median found in each bin with the error bars representing one standard deviation.  Solid black line is the least square fit to the median data points.  This fit is also representative of the average relationship at $z=6,7,8$, where $SFR \propto M_s^{1.10}$.  Blue squares and red circles show {\it HST/IRAC} detected objects at $z=7$ taken from \citet{Schaerer.deBarros:10,Labbe.etal:10}, while the blue triangles represent non {\it IRAC} detected objects at $z=7$ \citep{Schaerer.deBarros:10}.
}
\label{fig:ms_sfr}
\end{center}
\end{figure}

Fundamentally $\DCm$ and $\DCs$ are measure of the intrinsic scatter in the relationships between $M_s$ and $M_{UV}$ (Fig.~\ref{fig:ms_muv}), or between $M_s$ and $SFR$.  For each of the relationships, we find the scatter of about $\Delta \Muv \sim \pm 1$ mag and $\sim 1$ dex in SFR, respectively (Fig.~\ref{fig:ms_sfr}).  The duty cycle must be considered when estimating the intrinsic values from the observed results,  and when making comparisons between simulations and observations.  
 
In Fig. \ref{fig:ms_muv} we show the $M_s-\Muv$ relationship at $z=6$ with $\Ebv=0.10$ for N400L10 (yellow diamonds), N400L34 (yellow squares) and N600L100 (yellow triangles) runs, plotted with median points from observations by \citet[][blue squares]{Gonzalez.etal:10} and \citet[][red triangles]{Stark.etal:09} as well as $13$ spectroscopically confirmed $z=5.5-6.5$ galaxies presented in \citet[][black crosses]{CurtisLake.etal:12}.  The dashed blue line is a $z=4$ fit from the same \citet{Gonzalez.etal:10} work as above.  Here we find good agreement between observations and our simulations, although our median trend (solid black line) deviates from the $z=4$ Gonzalez relationship toward the low-mass end.  Further observations will be needed at lower masses to determine if this deviation is significant.

In Fig.~\ref{fig:ms_sfr} we demonstrate the scatter in the $M_s-SFR$ relationship with each of our runs at $z=7$, shown with observations from \citet{Labbe.etal:10} and \citet{Schaerer.deBarros:10}.  Agreement with observations is somewhat tentative as the scatter in the observed points is larger than in the simulated data.  This, in part, can be attributed to the debate as to whether or not nebular emission lines (NELs) need to be included in the SED fittings at high-$z$.  The \citet{Schaerer.deBarros:10} data points (blue) are obtained with NELs included, while the \citet{Labbe.etal:10} points (red) do not include NELs.


Currently there are inconsistencies between observations and theoretical predictions of the GSMF and SFRD.  Simulations predict steeper low-mass slopes of the GSMF ($\alpha_M$) than current observational estimates \citep{Gonzalez.etal:10} with a difference of $ \Delta \alpha_M \approx 0.80$ at $z=6$.  This difference is more significant than that of the faint-end slope of the UV LFs, which is $\Delta \alpha_L\approx 0.40$ \citep{Bouwens.etal:12, Jaacks.etal:12}.  
Likewise the theoretical predictions of SFRD at $z=6-10$ are in conflict with the estimates obtained by observations, showing an offset of more than one order of magnitude in total SFRD.
\citet{Choi.etal:12} showed that these discrepancies can be partly understood if we consider the mass and flux limits of the current high-$z$ galaxy surveys, however, they did not consider the effect of the duty cycle. 
In the following sections we show that, through the application of the $DC$, we can converge observations and theory to make a more consistent portrayal of the early Universe.

\subsection{Galaxy Stellar Mass Function}
\label{sec:GSMF}
\begin{figure*}
\begin{center}
\includegraphics[width=1.9\columnwidth,angle=0] {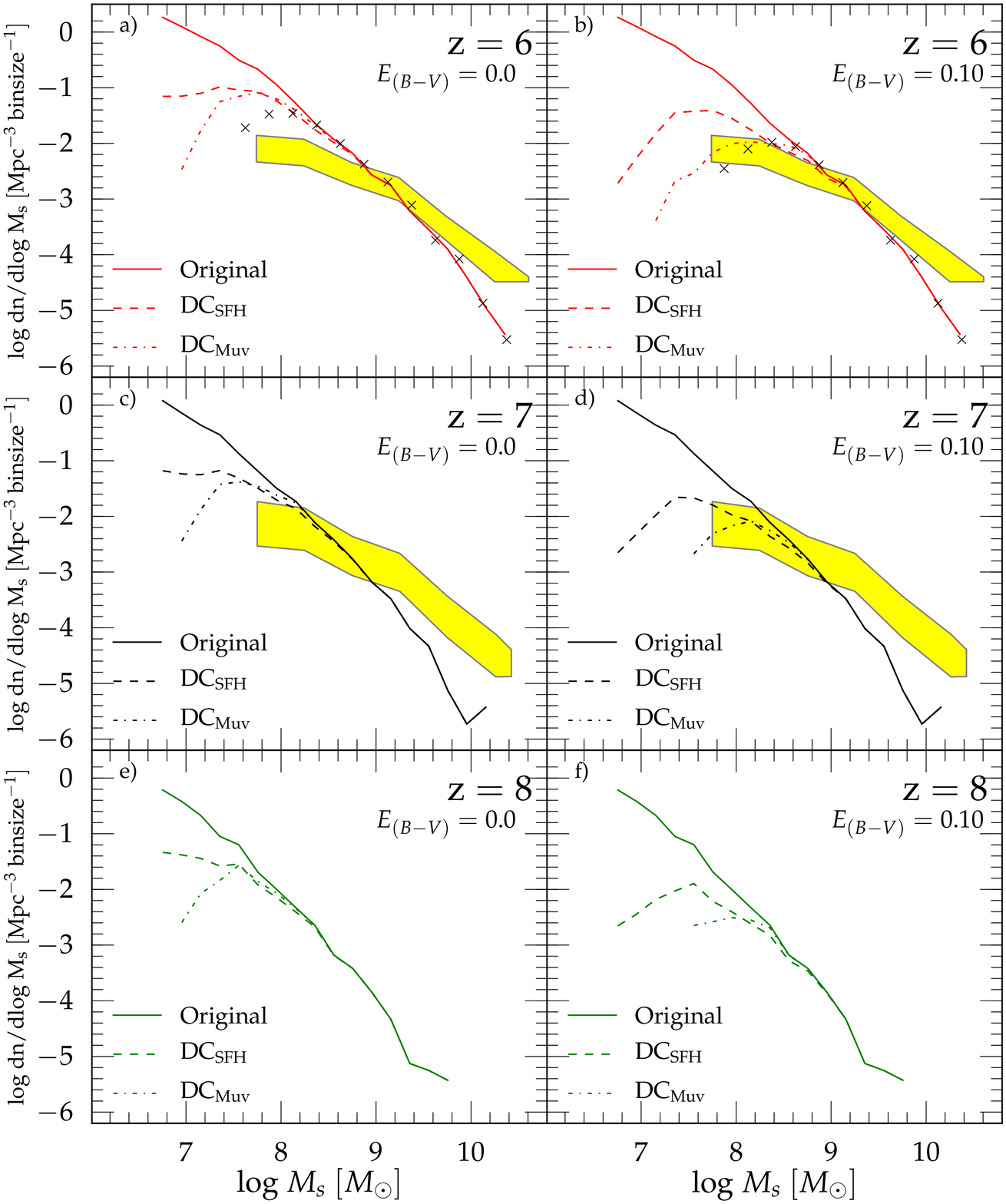}
\caption{GSMF (solid lines) at $z=6,7,8$ with both $\DCs$ (dashed lines) and $\DCm$ (dot-dashed line) applied.  This procedure is done with the effects of dust extinction (right column) and without (left column) at each redshift.  Yellow shade represents the GSMF derived from observations presented in \citet{Gonzalez.etal:10}. The crosses in panels a,b represent simple $\Muv$ cut of $\Muv=-18$ applied to the each run prior to computing the GSMF.
} 
\label{fig:mf}
\end{center}
\end{figure*}

The observed GSMF is estimated using a combination of the observed UV LF and $\Muv - M_s$ relation, both of which are quantities that have the $DC$ built in (i.e., what cannot be observed cannot be included).  
Our simulations give the intrinsic number of galaxies in each mass bin, and this does not take into account of the observable threshold in the rest-frame UV light. Therefore we have to apply the observational threshold captured by the $DC$ before a proper comparison with observations can be made. 
The application of $DC$ as a function of $M_s$ reduces the number of galaxies in certain mass bins where $DC<1$. 

In Fig.~\ref{fig:mf} we explore the effect of $\DCs$ (dashed line) and $\DCm$ (dot-dashed line) on the GSMF.  The solid lines in all panels represent the GSMF found in our simulations and originally presented in \citet{Jaacks.etal:12}.  The yellow shaded areas represent the GSMF derived from observations by \citet{Gonzalez.etal:10},  who utilize SED fits to obtain the $M_s - \Muv$ relationship and then a combination of the UV LF and the mass-to-light to determine the GSMF. We present these results with (right column) and without (left column) extinction correction as we described in Section~\ref{sec:duty}. 
These results are obtained through
\begin{equation}
\phi_{\rm obs}(z,M_s)=\phi_{\rm int}(z,M_s)\times DC(z,M_s),
\end{equation}
where $\phi_{\rm obs}(z,M_s)$ is the observed number density at a given redshift and stellar mass $M_s$, and $\phi_{\rm int}(z,M_s)$ is the value intrinsic to our simulations at the same redshift.

Taking the $DC$ effect into account results in the slope of the low-mass end of GSMF becoming shallower at all redshifts by $\Delta \alpha_M \approx0.26$, where $\Delta \alpha_M=|\alpha_M({\rm obs})-\alpha_M({\rm int})|$.  This decrease of $\alpha_M$ brings our GSMF into better agreement with observations at the low-mass end.  The difference seen between the left column ($\Ebv =0.0$) and right column ($\Ebv =0.10$) in Fig. \ref{fig:mf} can be attributed to the fact that the $DC$s calculated with dust extinction tend to transition from zero to unity in a mass range which is $\sim 0.50$ dex higher than those without. This is an expected effect, because as extinction increases less star formation will be visible, resulting in a lower duty cycle at a given mass.

In terms of faint-end slope reduction, the application of both $\DCm$ and $\DCs$ obtain similar results to the use of simple flux limit applied to our galaxy population, as seen by the black crosses in Fig.~\ref{fig:mf}.  Deviations between $\DCm$ and a simple flux limit can be attributed to the fact that the the DC is a fit to a scatter in a particular mass bin (error bars in 
Fig.~\ref{fig:duty}), thereby allowing for the inclusion of more objects which would otherwise be excluded by a simple cut.


\subsection{Star Formation Rate Density (SFRD)}
\label{sec:sfrd}

\begin{figure}
\centerline{\includegraphics[width=1.10\columnwidth,angle=0] {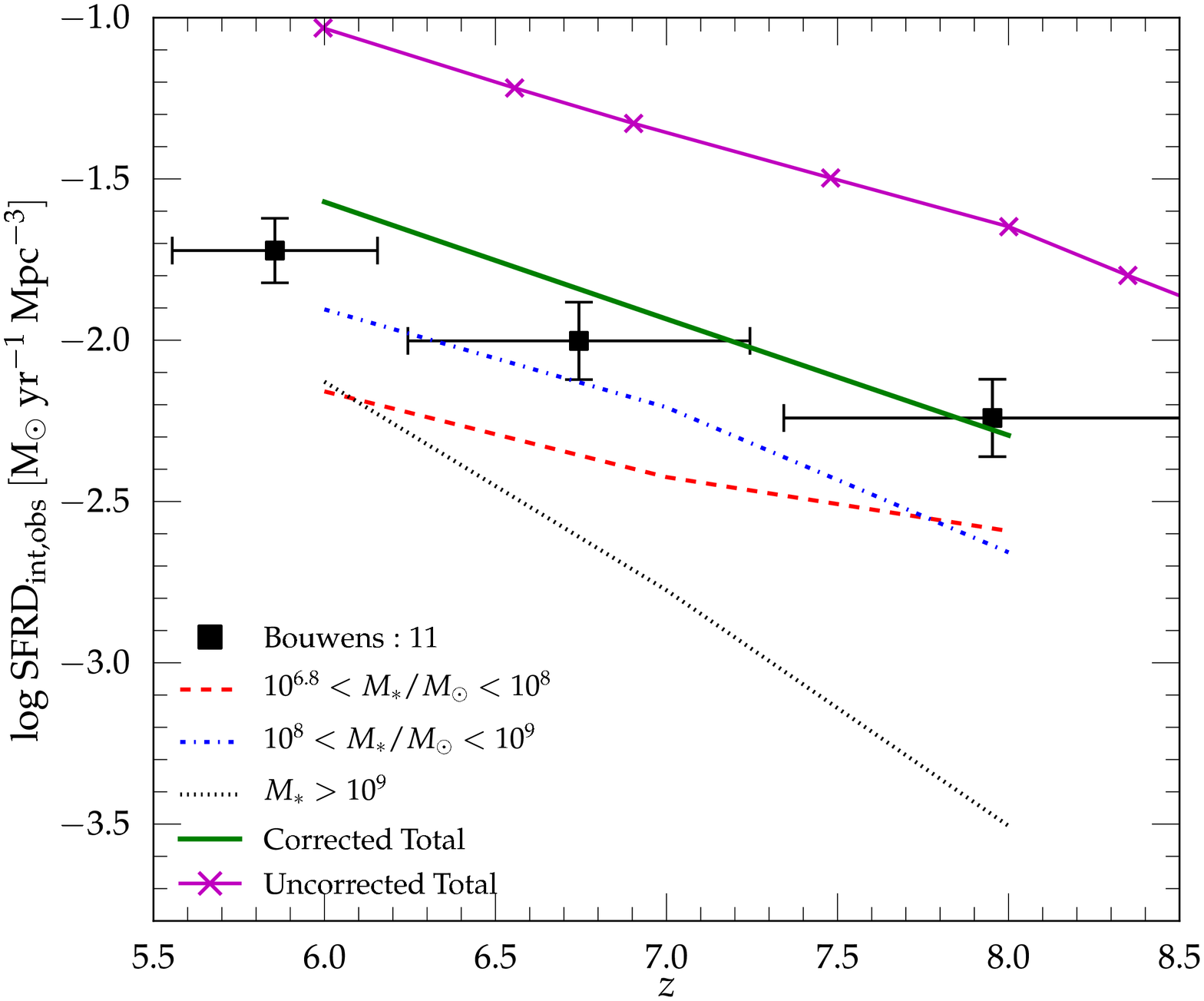}}
\caption{
Star formation rate density (SFRD), both intrinsic (SFRD$_{\rm int}$) and derived observable (SFRD$_{\rm obs}$), as a function of redshift shown with $\DCs$ applied.  The result for different galaxy stellar mass ranges are shown: $10^{6.8}<M_s/\Msun<10^{8}$ (dashed red), $10^{8}<M_s/\Msun<10^{9}$ (dash-dot blue) and $M_s>10^{9} \Msun$ (dotted black).  The total observable amount of all mass ranges (SFRD$_{obs})$ is shown as the dark green solid line.  Total SFRD$_{int}$ without $\DCs$ applied (i.e. the true total in our simulation) is shown by solid magenta line with crosses. Observational estimates taken from \citet{Bouwens.etal:12} (black squares). 
See \citet{Jaacks.etal:12} for details regarding justification of the mass range taken from each run. 
}
\label{fig:sfrd}
\end{figure}

The SFRD is also highly effected by the $DC$.  The observed SFRD is obtained by integrating the UV LF to the observed flux limits ($\Muv \sim -18$ at $z=6$) and converted to SFRD via the standard conversion factor depending on the IMF.  
Therefore if we wish to compare our simulated SFRD to the observed SFRD, we must take the $DC$ into account.  For example, if only a fraction of galaxies of a given $M_s$ are observable, then only that fraction can be allowed to contribute to the total SFRD at a particular redshift.  In principle this is similar to applying a flux limit to our simulated galaxies and the exclusion of those which do not reach an observable limit, and the same effect can be taken care of by the $DC$. 

Fig.~\ref{fig:sfrd} demonstrates the effect of applying the $\DCs$ with $\Ebv = 0.10$ to the SFRD at $z=6,7,8$.  By breaking down the galaxy population at each redshift into mass bins of $10^{6.8}<M_s/\Msun<10^{8}$ (dashed red), $10^{8}<M_s/\Msun<10^{9}$ (dot-dashed blue) and $M_s>10^{9} \Msun$ (dotted black), we are able to examine the contribution of each to the total SFRD.  This also allows us to apply our $\DCs$ to each of galaxies contained in these mass bins and remove those galaxies which are not currently observable from our total simulated galaxy population.  To account for the rapid transition in the $\DCs$ and the large bin size of the decomposed SFRD, we apply the $\DCs$ at $0.10\ \Msun$ intervals using
\begin{equation}
SFRD_{\rm obs}(z,M_s)=SFRD_{\rm int}(z,M_s)\times DC(z,M_s),
\end{equation}  
where $SFRD_{\rm obs}(z,M_s)$ is what would be observable at a particular redshift, and $SFRD_{\rm int}(z,M_s)$ is the intrinsic value in our simulation at the same redshift.
Although we didn't indicate explicitly in the equation above, note that $DC$ is also dependent on the values of limiting magnitude of the survey. 

This implementation results in decreasing the summed total SFRD (solid magenta line with crosses) found in our previous work \citep{Jaacks.etal:12} by $71\%,73\%$ and $77\%$ at $z=6,7,8$, respectively, as shown by the solid green line.  
The most recent observational data is shown by the black squares from \citet{Bouwens.etal:12}.  
Our results suggest that the $DC$ of galaxies could reduce the observable fraction of SFRD down to only $20-30$\%, and it is necessary to correct for its effect in order to estimate the observable fraction of total SFRD in the Universe.


\section{ Conclusions \& Discussions}
\label{sec:con}

\subsection{Main Conclusions}

Using cosmological SPH simulations, we examined the form of SFH of galaxies at $z\ge 6$, the duty cycle of star formation and its impact on the current observations.  Our major conclusions are as follows: 

\begin{itemize}
\item We find that, on average, the simulated SFH has an increasing form as a function of cosmic time, which can be characterized well by a power-law and an exponential function (Fig.~\ref{fig:sfh6}).  
The best-fitting characteristic times are summarized in Tables~\ref{tbl:fit_exp} and \ref{tbl:fit_power} for different stellar mass range of galaxies and have characteristic time-scales of 70\,Myr to 200\,Myr for galaxies with stellar masses $M_s \sim 10^6 \Msun$ to $>10^{10} \Msun$.  
We find that the power-law produces on average slightly better fit to the mean SFH of galaxies, but the exponential form also works quite well. 

\vspace{0.2cm}
\item The SFH of individual galaxies have bursts of star formation at $z\ge 6$, and these burst lead to a scatter in the relationship between  $M_s$ and $\Muv$ of $\Delta \Muv \sim \pm1$ mag and a $\sim 1$ dex scatter in the $M_s - SFR$ relation (Fig. \ref{fig:muv_sfr}).  The measured values of duty cycle reflect the scatter in these relationships.  \citet{Feldmann.etal:12} also supports the idea that star formation is inherently a stochastic process as they have implemented a model treating SF as a Poisson process within GMCs.  \citet{Forero-Romero.etal:12} have shown that SF stochasticity can broaden the distribution of Ly$\alpha$ equivalent width, and impact the observed scatter in the $\Muv-SFR$ relationship of Ly$\alpha$ emitters with $SFR\leq 10^{-2}\Msunr$.   These work support the idea that 
duty cycle can be a useful quantity to summarize the scatter in the $M_s - \Muv - SFR$ relationships.

While \citet{Finlator.etal:11} find a very similar mean $M_s - SFR$  relationship in their simulations they find a much smaller scatter which is  on the order of ~0.2 dex using comparable resolution runs.  They interpret this, along with their smoothly rising SFH, as evidence for smooth accretion rather than merger-driven growth. In contrast, our $\sim 1$ dex scatter suggests that we can not rule out the dominance of merger-driven growth for high-z galaxies, especially at $z\ge 6$ when the merger rates are very high and gravitational growth of structure is driving the rapid increase in cosmic SFRD.

\vspace{0.2cm}
\item  We measured the duty cycle of simulated galaxies using two different methods: 1) $\DCs$, based on the galaxy's star formation history; 2) $\DCm$, based on the instantaneous UV magnitude of the galaxy. Both methods give $DCs$ at $z=6,7,8$ that make a steep transition from zero to unity at  $M_s = 10^7 - 10^9\,M_\odot$.  
The effect of dust extinction comes in when we convert the magnitude limit into threshold SFR ($SFR_{\rm th}$) using the relation between the two quantities in our simulations. 
The range in which the transition occurs is highly dependent on the amount of dust extinction assumed to be present.  With the dust effect of $E_{B-V}=0.10$, the transition mass-scale is about $\log (M_s/\Msun) = 8.1-8.4$, but without the dust effect, the transition occurs at $\log (M_s/\Msun) = 7.6 - 7.9$ (see Table~\ref{tbl:50duty} and \ref{tbl:fit_duty}). This may be better characterized as a shift to a lower $DC$ for a given $M_s$ as fewer galaxies will be able to exceed either the SFH or $\Muv$ thresholds as $\Ebv$ increases. 
With a greater value of $E_{B-V}$, the galaxy has to have a higher SFR (and hence higher $M_s$) in order to exceed the observable threshold. 

\vspace{0.2cm}
\item The application of $\DCs$ including $E_{B-V}=0.10$ to the GSMF results in lowering the faint-end of the GSMF by on average $\Delta \alpha \approx 0.26$ at $z=6,7,8$ (Fig. \ref{fig:mf}), which is similar to the results found at lower redshifts by \citet{Lee.etal:11}.

\vspace{0.2cm}
\item  The application of the duty cycle to the simulated SFRD results in lowering the total SFRD by $71, 72, 77\%$ at $z=6, 7, 8$, respectively (Fig.~\ref{fig:sfrd}).  This indicates that the current observations (even the deepest HST galaxy imaging surveys probing down to $\Muv \sim -18$ mag)  could be missing a significant fraction ($\gtrsim 70$\%) of star formation in the early Universe.

\end{itemize}


\subsection{$\rm H_2$ regulated star formation}
The details of the SFHs of simulated galaxies could be dependent on the SF model implemented in our simulations.  
Therefore the discrepancies that we find between our results and that of \citet{Finlator.etal:11} may simply be due to the differences between the adopted SF model. 
Currently we use the "Pressure model" \citep{Schaye:08, Choi:10a}, in which stars are allowed to form once the gas density exceeds the threshold density of $n_{\rm th}^{\rm SF}=0.6$\,cm$^{-3}$ \citep[see][for the justification of this value]{Nagamine.etal:10}. Star particles are then generated based on the gas pressure of star-forming gas according to the SF law matched to the local \citet{Kennicutt:98} relationship.

There is observational evidence to suggest that star formation in the Universe is controlled by the fraction of H$_2$ \citep{Kennicutt.etal:07, Leroy.etal:08, Bigiel.etal:08}. 
This could lead to a reduction in star formation at high redshift due to lower metallicity \citep{Krumholz.etal:09,Kuhlen.etal:12}, and would alter the SFHs in our simulation.  Therefore this work would need to be revisited with a new SF model based on H$_2$ mass. 
We are currently in the process of implementing a new SF model based on H$_2$ mass based on the work by \citep{Krumholz.etal:09} in our simulations, and we will report the results in a subsequent paper (Thompson, et al., in preparation).


\section*{Acknowledgments}

We are grateful to V. Springel for allowing us to use the original version of {\small GADGET-3} code, on which the \citet{Choi:10a, Choi.etal:12} simulations are based.  We would like to thank Stephen Wilkins, Michele Trenti and Richard Bouwens for useful discussions and the referee for the constructive feedback.  JJ is partially supported by the Nevada NASA EPSCoR Cooperative Agreement NNX07AM20A and the Nevada System of Higher Education.  This work was supported in part by the NSF grant AST-0807491, National Aeronautics and Space Administration under Grant/Cooperative Agreement No. NNX08AE57A issued by the Nevada NASA EPSCoR program, and the President's Infrastructure Award from UNLV. 
Support for Program number HST-AR-12143-01-A was provided by NASA through a grant from the Space Telescope Science Institute, which is operated by the Association of Universities for Research in Astronomy, Incorporated, under NASA contract NAS5-26555.  This research is also supported by the NSF through the TeraGrid resources provided by the Texas Advanced Computing Center (TACC) and the National Institute for Computational Sciences (NICS). Some numerical simulations and analyses have also been performed on the UNLV Cosmology Cluster.  
KN acknowledges the hospitality and the partial support from the Kavli Institute for Physics and Mathematics of the Universe (IPMU), University of Tokyo, the Aspen Center for Physics, and the National Science Foundation Grant No. 1066293.


\bibliographystyle{mn2e}

\end{document}